\documentclass{article}

\usepackage[preprint]{neurips_2026}
\usepackage[utf8]{inputenc} 
\usepackage[T1]{fontenc}    
\usepackage{hyperref}       
\usepackage{url}            
\usepackage{booktabs}       
\usepackage{amsfonts}       
\usepackage{nicefrac}       
\usepackage{microtype}      
\usepackage{xcolor}    
\usepackage{amsmath}
\usepackage{bm}
\usepackage{float}
\usepackage{tikz}
\usepackage{wrapfig}
\usepackage{listings}
\usetikzlibrary{shapes.geometric, positioning, calc, arrows.meta}
\definecolor{scnb_blue}{RGB}{197, 224, 240}
\definecolor{en_head_orange}{RGB}{247, 220, 180}
\definecolor{ch_head_green}{RGB}{200, 230, 201}
\definecolor{fo_head_purple}{RGB}{216, 202, 233}

\tikzset{
    block/.style={
        draw, thick, rounded corners=12pt, align=center,
        minimum width=4.5cm, minimum height=3.0cm, inner sep=12pt, fill=scnb_blue
    },
    headblock/.style={
        draw, thick, rounded corners=10pt, align=center,
        minimum width=4.2cm, minimum height=1.8cm, inner sep=10pt
    },
    arrow/.style={->, >=stealth, thick},
}

\title{Multitask learning with semiempirical orbital charges enables sample-efficient MLIPs}

\vspace{-6pt}
\author{%
  Ihor Neporozhnii \\
  Department of Physical and Environmental Sciences \\
  University of Toronto \\
  Toronto, ON M1C 1A4, Canada \\
  The Alliance for AI-Accelerated Materials Discovery \\
  Toronto, ON M5S 3G4, Canada \\
  \texttt{ihor.neporozhnii@mail.utoronto.ca} \\
  \And
  Sjoerd Hoogland \\
  Department of Electrical and Computer Engineering \\
  University of Toronto \\
  Toronto, ON M5S 3G4, Canada \\
  The Alliance for AI-Accelerated Materials Discovery \\
  Toronto, ON M5S 3G4, Canada \\
  \texttt{sjoerd.hoogland@utoronto.ca} \\
  \And
  Oleksandr Voznyy \\
  Department of Physical and Environmental Sciences \\
  University of Toronto \\
  Toronto, ON M1C 1A4, Canada \\
  The Alliance for AI-Accelerated Materials Discovery \\
  Toronto, ON M5S 3G4, Canada \\
  \texttt{o.voznyy@utoronto.ca} \\
}

\begin{document}

\maketitle

\vspace{-11pt}
\begin{abstract}
\vspace{-5pt}
  Machine learning interatomic potentials (MLIPs) require generating computationally expensive, large-scale training datasets to accurately simulate materials and molecules. Incorporating electronic structure information using multitask learning improves sample efficiency, however, training on full Hamiltonian matrices, which scale quadratically with the number of atoms, is intractable for large datasets. In this work, we show that multitask learning utilizing orbitally resolved semiempirical charges significantly improves sample efficiency and accuracy in MLIPs. To efficiently predict orbital charges, we implement a specialized equivariant model, reducing charge prediction error compared to an invariant baseline. By augmenting training with computationally inexpensive GFN1-xTB orbital charges, which scale linearly with the number of atoms, our model achieves a 58\% reduction
in energy mean absolute error relative to energy-only baselines, and requires five times less data to match their performance. Against the stronger baseline of joint energy-and-force training, orbital charges still reduce the energy error by 12\%. Furthermore, our approach matches or outperforms models trained on expensive density functional theory (DFT) atomic charges, capturing orbitally resolved electronic complexity and forcing the network to learn a physically accurate latent space that spontaneously clusters metals by shared chemical properties. Because orbital charges are only required during training, this approach preserves inference efficiency, providing a scalable recipe for developing accurate, data-efficient foundation models for complex chemical systems.
\end{abstract}

\section{Introduction}

MLIPs have fundamentally transformed the landscape of computational chemistry, enabling the simulation of complex systems at length and time scales previously inaccessible to first-principles methods. By learning the underlying quantum mechanical interactions, these models are accelerating discoveries across diverse domains, from solid-state materials to drug design \cite{gasteiger_gemnet_2021, musaelian_learning_2023, batatia_mace_2023, wood_uma_2026}. However, the accuracy and transferability of MLIPs are bound by the size and quality of their training datasets. Generating this reference data using high-fidelity electronic structure methods, such as density functional theory (DFT), requires immense computational resources. For instance, constructing the recent OMol25 dataset required approximately 6.6 billion CPU hours \cite{levine_open_2026}. To mitigate this data bottleneck, multitask learning and transfer learning have emerged as powerful strategies. By training MLIPs to predict additional electronic properties, models can achieve higher sample efficiency. Recent work has demonstrated that training on the full electronic Hamiltonian drastically improves MLIP performance on the energy prediction task \cite{kaniselvan_learning_2025}. Yet, because Hamiltonian matrices scale quadratically $O(N^2)$ with the number of atoms, computing, storing, and training on these massive objects is computationally intractable for large-scale or high-throughput applications.

A natural proxy for electronic structure is the distribution of charge. Due to the historical scarcity of charge labels, MLIPs have traditionally incorporated atomic charge information through the inclusion of local magnetic moments \cite{deng_chgnet_2023} or via classical charge equilibration schemes \cite{ghasemi_interatomic_2015, ko_fourth-generation_2021}. While equilibration methods with quasi-linear scaling have been developed \cite{gubler_accelerating_2024}, they frequently suffer from delocalization and overpolarization errors \cite{vondrak_pushing_2025}. Multitask learning frameworks that use atomic charges have recently achieved state-of-the-art performance on purely organic systems \cite{zubatyuk_accurate_2019, anstine_aimnet2_2025}. Extending these methods to complex organometallic molecules, which are critical for catalysis, atomic layer deposition, and optoelectronics, remains challenging. In transition metals, the directional nature of $d$-orbitals dictates chemical reactivity, meaning that atomic charges lack the spatial complexity of the electronic structure.

In this work, we demonstrate that multitask learning utilizing orbitally resolved semiempirical charges significantly improves the sample efficiency and accuracy of current MLIPs. We augment our training data with orbital charges derived from the computationally inexpensive GFN1-xTB tight-binding method \cite{grimme_robust_2017}, which, unlike full Hamiltonians, scales linearly $O(N)$ with system size. To effectively learn these properties, we introduce an orbital charge-adapted equivariance framework derived from the transformation rules of density matrices under rotations. This architecture introduces negligible computational overhead compared to energy-only baselines while properly embedding the physical symmetries of atomic orbitals into the network.

We show that this orbital-multitask approach reduces the energy mean absolute error by 58\%
compared to energy-only baselines at our largest training set size, requiring five times less
training data to achieve equivalent performance. Crucially, we demonstrate that incorporating
inexpensive, orbitally resolved GFN1-xTB charges provides an inductive bias at least as strong as
training on computationally expensive DFT atomic charges, and that it is the best-performing charge
target at every dataset size we examined. By capturing orbital-level electronic complexity, our
network learns a chemically rigorous latent space that spontaneously clusters elements by their
chemical properties. Because these orbital charges are required only during training, our framework
preserves efficient inference speeds of standard MLIPs, establishing a highly scalable and physically
grounded approach for the next generation of chemical foundation models.

\vspace{-5pt}
\section{Results}
\vspace{-5pt}
\subsection{Multitask Performance and Sample Efficiency}
\vspace{-5pt}

\begin{figure}[H]
    \centering
    \begin{tikzpicture}
        \tikzset{
            block/.style={
                draw, thick, rounded corners=5pt, align=center,
                minimum width=1.6cm, minimum height=3.2cm, inner sep=5pt
            },
            headblock/.style={
                draw, thick, rounded corners=5pt, align=center,
                minimum width=2.2cm, minimum height=0.9cm, inner sep=5pt
            },
            arrow/.style={->, >=stealth, thick},
        }

        \node[anchor=north west, inner sep=0] (fullfig) at (0,0) {
            \includegraphics[width=\textwidth]{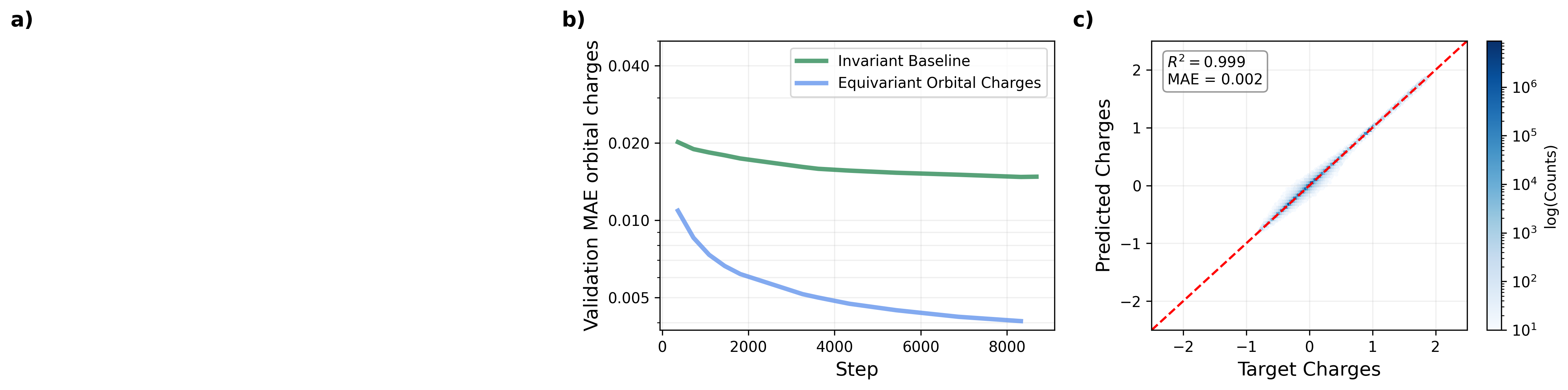}
        };

        \begin{scope}[scale=0.65, transform shape,
                      shift={($(fullfig.north west)+(0.085\textwidth,-0.19\textwidth)$)}]]

            \node[align=center, font=\footnotesize, text width=1.5cm] (geom_label)
                {\textbf{Molecular} \\ \textbf{Geometry}};

            \node[block, fill=scnb_blue, right=0.3cm of geom_label.east, font=\footnotesize]
                (backbond) {\textbf{eSCN}\\\textbf{Backbone}};
            \draw[arrow] (geom_label.east) -- (backbond.west);

            \coordinate (split_e) at ($(backbond.south east)!0.8333!(backbond.north east)$);
            \coordinate (split_f) at ($(backbond.south east)!0.5!(backbond.north east)$);
            \coordinate (split_q) at ($(backbond.south east)!0.1667!(backbond.north east)$);

            \node[headblock, fill=en_head_orange, right=0.5cm of split_e, font=\footnotesize]
                (energy_head) {\textbf{Energy Head}};
            \node[headblock, fill=fo_head_purple, right=0.5cm of split_f, font=\footnotesize]
                (force_head)  {\textbf{Forces Head}};
            \node[headblock, fill=ch_head_green, right=0.5cm of split_q, font=\footnotesize]
                (charge_head) {\textbf{Charge Head}};

            \draw[arrow] (backbond.east |- split_e) -- (energy_head.west);
            \draw[arrow] (backbond.east |- split_f) -- (force_head.west);
            \draw[arrow] (backbond.east |- split_q) -- (charge_head.west);

            \draw[arrow] (energy_head.east) -- ++(0.3cm, 0) node[right, font=\normalsize] {$E$};
            \draw[arrow] (force_head.east)  -- ++(0.3cm, 0) node[right, font=\normalsize] {$\bm{F}_a$};
            \draw[arrow] (charge_head.east) -- ++(0.3cm, 0) node[right, font=\normalsize] {$\bm{q}_a$};
        \end{scope}
    \end{tikzpicture}

    \vspace{-10pt}
    \caption{Model architecture and equivariant orbital charge prediction. a) Schematic of the multitask model for predicting total energy, atomic forces, and orbital charges on each atom using the eSCN backbone. The molecular geometry is passed through the shared backbone, splitting into standard energy and force heads and an equivariant orbital charge head. b) Validation Mean Absolute Error (MAE) for orbital charge prediction as a function of training steps. The equivariant orbital charge architecture demonstrates a significantly reduced error compared to the invariant baseline. c) Parity plot of predicted versus target orbital charges.}
    \label{fig:architecture_and_eqq}
    \vspace{-20pt}
\end{figure}

In molecular machine learning, equivariant representations play a key role in modeling underlying physical symmetries \citep{Thomas_2018, Batzner_2022}. To simultaneously predict total energies and orbital charges, we developed a multitask architecture with separate prediction heads built upon a shared equivariant Spherical Channel Network (eSCN) \cite{pmlr-v202-passaro23a} backbone. By passing the atomic embedding vectors through a shallow, equivariant charge head, our model predicts orbitally resolved charges consistent with the GFN1-xTB basis set with minimal computational overhead (see Section \ref{sec4} for details). Incorporating how orbital charges transform under molecular rotations directly into the charge head allows the network to preserve the spatial symmetries of atomic orbitals, reducing the charge prediction error by a factor of more than four relative to an invariant baseline model (Fig. \ref{fig:architecture_and_eqq}b).

\vspace{-10pt}

\begin{figure}[H]
    \centering
    \includegraphics[width=1.0\textwidth]{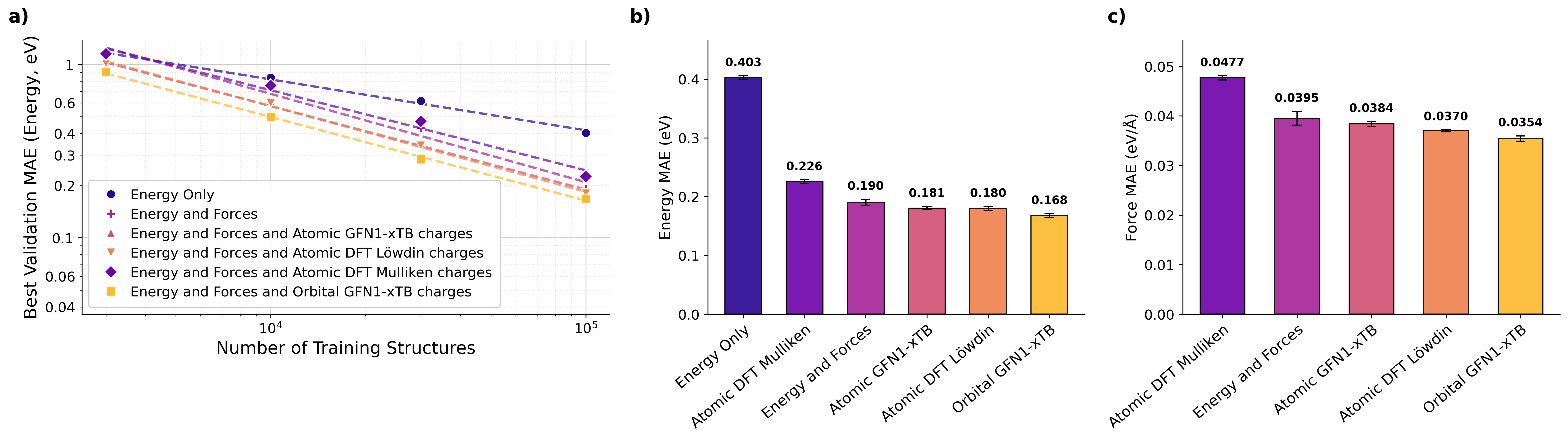}
    \centering
    \vspace{-15pt}
    \caption{Dataset scaling and sample efficiency. a) Log-log plot of the best validation energy MAE as
a function of the number of training structures, for all six training settings. Multitask training with
GFN1-xTB orbital charges consistently outperforms energy-only training, energy-and-force training, and
the atomic-charge multitask baselines across all data regimes, achieving equivalent error with five
times less data than the energy-only baseline. b) Energy MAE and c) force MAE at 100{,}000 training
structures. Bars show the mean over three independent seeds with error bars representing standard deviation.}
    \label{fig:orbital_charges}
\end{figure}

The primary objective of MLIP models is to learn the potential energy surface, enabling the prediction of dissociation energies, conformational landscapes, and atomic forces (negative gradients of the energy with respect to atomic positions), which in turn enables molecular dynamics simulations. Accurate energy prediction is therefore essential for stable and reliable MLIPs. Notably, multitask learning with semiempirical orbital charges substantially improves the accuracy of DFT total energy predictions. Our orbital-charge multitask framework reduces the energy mean absolute error (MAE) by 58\% relative to the baseline trained solely on DFT energy labels, and by 12\% relative to the model trained jointly on energy and force labels (Fig.~\ref{fig:orbital_charges}).

\vspace{-7pt}

To isolate and evaluate the distinct advantage of orbitally resolved data, we benchmarked our framework against models trained on standard atomic charges: semiempirical GFN1-xTB atomic charges, alongside DFT Löwdin and Mulliken atomic charges from the OMol25 dataset. These target labels present fundamentally different representations of electron distribution. For example, in the presented cadmium complex, GFN1-xTB provides highly localized atomic charges, placing negative charge primarily at oxygen or nitrogen atoms while keeping the carbon backbone largely neutral. The Löwdin method introduces broader, moderate polarization across the ligand, whereas the Mulliken analysis exhibits extreme charge separation characterized by heavily exaggerated dipoles.

The network reproduces GFN1-xTB and DFT Löwdin atomic charges almost exactly (charge MAE
0.002 e, see Appendix Fig.~\ref{fig:atomic_charge_prediction}), but not the highly polarized Mulliken charges (MAE 0.076 e). These differences carry into energy prediction performance: GFN1-xTB and Löwdin atomic charge supervision result in MAE of 0.181 and 0.180 eV, both improving modestly on energy-and-force training, while training on atomic Mulliken charges is harmful for both energy and force tasks. This suggests that selecting an appropriate charge partitioning approach is crucial for achieving improvements on adjacent training objectives.

Beyond improved accuracy, our multitask strategy yields exceptional gains in sample efficiency. The multitask model requires five times less training data to match the energy prediction accuracy of the energy-only baseline (Fig.~\ref{fig:orbital_charges}). While this approach was evaluated in a relatively low-data regime with the largest training set size of $10^5$ molecules, an analysis of the dataset size learning curves suggests even greater potential. 

\vspace{-7pt}
\subsection{Emergent Chemistry in the Latent Space}
\vspace{-3pt}

Ensuring that models learn latent representations that capture accurate physical and chemical information is critical for achieving robustness and generalizability. By analyzing and understanding these representations, we can evaluate performance, detect issues, and identify opportunities for improvement \citep{edamadaka2025universally, li2026platonic}. To understand how the choice of training labels impacts the learned representation of atomic environments, we visualized the backbone atomic embeddings using Uniform Manifold Approximation and Projection (UMAP) \citep{McInnes2018}. Because of the dataset's focus on organometallic complexes, we concentrated this structural analysis specifically on the embeddings of transition metal elements.  

\begin{figure}[H]
    \centering
    \includegraphics[width=1.0\textwidth]{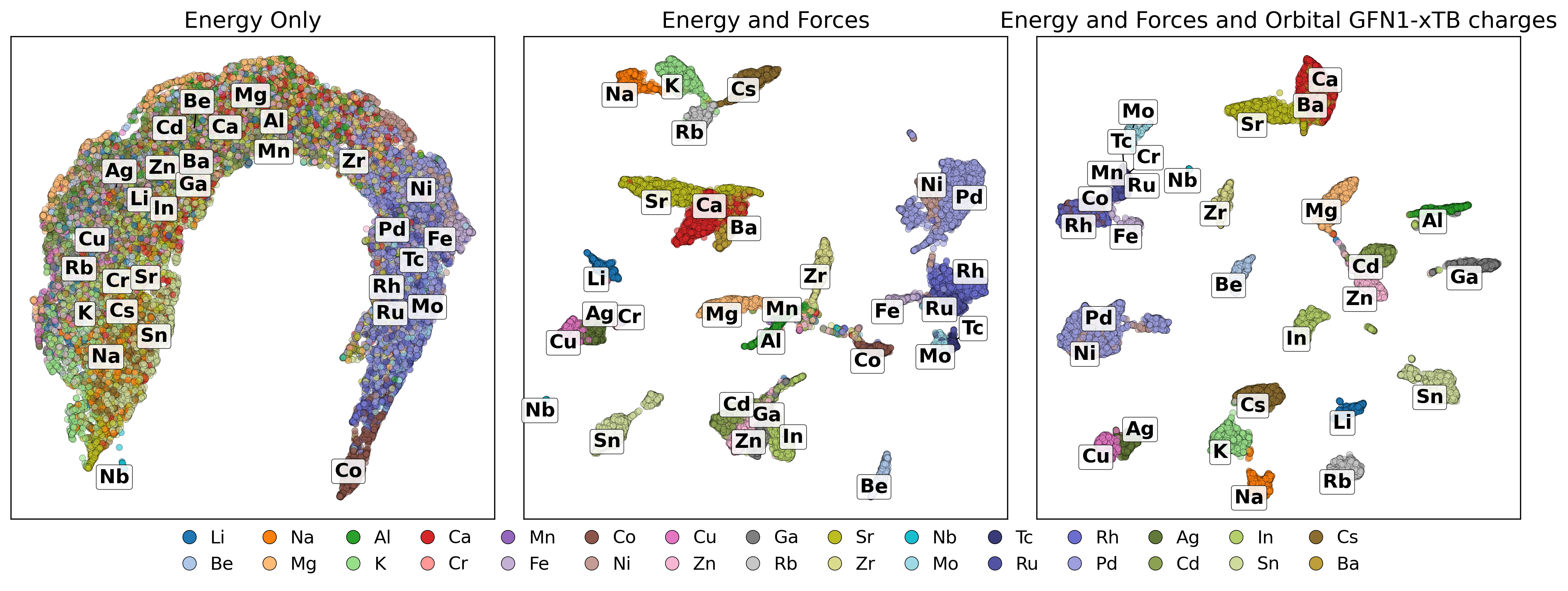}
    
    \caption{UMAP projections of transition metal atomic embeddings extracted from the eSCN backbone after training. Panels compare the representations learned via energy-only training against multitask training using various charge types. Multitask training with GFN1-xTB orbital charges forces the network to learn a chemically rigorous latent space, spontaneously clustering elements that share similar chemical behavior (e.g., Ag/Cu, Zn/Cd, Ca/Ba, Ni/Pd).}
    \label{fig:umap}
\end{figure}

Our latent space projections demonstrate that multitask training on energies paired with semiempirical orbital charges induces a highly ordered, physically grounded atomic representation. Driven by the orbital charge data, the network spontaneously clusters elements with shared chemical and physical properties, such as Ag/Cu, Zn/Cd, Ca/Ba, and Ni/Pd (Fig. \ref{fig:umap}). In stark contrast, models trained exclusively on energies fail to organize atomic environments into these chemically meaningful clusters. The emergence of this chemically rigorous latent space provides a fundamental explanation for the substantial performance improvements and enhanced generalizability realized by our orbital-charge multitask framework.  

\vspace{-20pt}

\section{Methods}\label{sec4}
\vspace{-5pt}
\subsection{Equivariant prediction of orbital charges}\label{subsec41}
\vspace{-5pt}

The orbital populations $\bm\nu$ and charges $\mathbf{q}$ are computed from the density and overlap matrices, as well as the reference occupations $\bm\nu_0$ using element-wise multiplication: 
\begin{align}
    \nu_i = \sum_{j=1}^{N_{orb}} (\mathbf{P} \circ \mathbf{S})_{ij} = \sum_{j=1}^{N_{orb}} P_{ij}S_{ij} \quad \text{and} \quad \mathbf{q} = \mathbf{\nu_0} - \mathbf{\nu}
\end{align}

The transformation of the density matrix $\mathbf{P}$ and overlap matrix $\mathbf{S}$ under molecular coordinate rotations is captured by the rotation matrix $\mathbf{R}$ and the corresponding Wigner-D matrix {$\mathcal{D}(\mathbf{R})$}: 

\begin{equation}
        \mathbf{P'} = \mathcal{D} \mathbf{P} \mathcal{D}^T \quad \text{and} \quad \mathbf{S'} = \mathcal{D} \mathbf{S} \mathcal{D}^T
\end{equation}

We show that exact orbital charge transformations under molecular rotations are captured by Equation (\ref{eq:3}), where $\mathcal{B}(i)$ represents the angular momentum block of $i^{th}$ orbital (see Appendix \ref{secB} for the complete derivation of equivariant orbital transformation rules).

\begin{equation}
        \nu'_i = \sum_{k,m \in \mathcal{B}(i)} \mathcal{D}_{ik} (\mathbf{P}\mathbf{S})_{km} \mathcal{D}_{im}
        \label{eq:3}
\end{equation}

From Equation (\ref{eq:3}), it follows that orbital charges depend only on the block-diagonal elements of the density and overlap matrices. This allows us to predict them efficiently without storing or reconstructing the full matrices, achieving linear scaling $O(N)$ with system size.

To implement orbital charge prediction, we adopt the equivariant eSCN backbone. This architecture has previously demonstrated strong performance in predicting the charge densities of small organic molecules by implicitly learning molecular wavefunctions, a task closely related to orbital charge prediction \citep{fu2024recipe}. The eSCN architecture uses spherical harmonics irreducible representations (irreps) to enforce $SO(3)$ equivariance, making it naturally aligned with the angular structure of atomic orbitals and therefore well suited for orbital-level prediction tasks. The backbone model $\mathcal{F}_{[\Theta]}$ requires only atomic types $\mathcal{A}$ and atomic position $\mathcal{R}$ as input. After the final convolutional block, each atom $a \in N$ is represented by an embedding vector $x_a$ consisting of 128 irreps for each angular momentum level $L$, corresponding to orbital-like features.

The atomic embedding $x_a$ is then processed by three prediction heads: an invariant head
$f_{E[\theta_E]}$, which predicts the total molecular energy, an equivariant head $f_{F[\theta_F]}$,
which predicts atomic forces, and an equivariant head $f_{q[\theta_q]}$, which predicts orbitally
resolved charges for atom $a$ consistent with the GFN1-xTB basis:
\vspace{-15pt}

\begin{gather}
 \{x_a, a \in N \} = \mathcal{F}_{[\Theta]}(\mathcal{A}, \mathcal{R})  \\
  E_a = f_{E[\theta_E]}(x_a) \quad \text{and} \quad \mathbf{F}_a = f_{F[\theta_F]}(x_a) \quad \text{and} \quad \mathbf{q}_a = f_{q[\theta_q]}(x_a)
\end{gather}

\subsection{Data generation and processing}\label{main_data}

To construct the training dataset, we curated a subset of neutral organometallic complexes, evaluated
in a singlet state (spin multiplicity 1), sampled from the Train subset of the OMol25 4M dataset. This
provided the 3D molecular structures, DFT total energies, DFT forces, and DFT Mulliken and L\"owdin atomic charges.
After filtering (Appendix~\ref{secC}), the curated dataset comprises 160{,}000 molecules in total,
partitioned into a training pool of 100{,}000 structures, a validation set of 10{,}000, and a test set
of 50{,}000. Models were trained on nested subsets of 3{,}000, 10{,}000, 30{,}000, and 100{,}000
structures drawn from the training pool, while the same validation and test sets were held fixed across
every model and run, so that all reported metrics are directly comparable.

Semiempirical GFN1-xTB orbital charges were computed with the DXTB package~\cite{friede_dxtb_2024}
(0.7~s per molecule on a single CPU core) and filtered for convergence reliability. The atomic charges
were obtained by summing orbital charges per atom. Full data-generation and filtering details are
given in Appendix~\ref{secC}.

\vspace{-5pt}
\section{Discussion}
\vspace{-5pt}

Machine learning interatomic potentials have become indispensable tools for accelerating discovery across the chemical sciences, driving advancements from solid-state materials development to targeted drug design. However, the predictive accuracy and generalizability of these models are fundamentally constrained by the availability of high-fidelity training data. Generating this reference data utilizing ab initio methods like DFT is exceptionally demanding; for example, constructing the recent OMol25 dataset required billions of CPU hours. As frontier datasets surpass hundreds of millions of structures, standard data generation strategies are beginning to face diminishing returns, requiring orders of magnitude more computational power to achieve only marginal performance improvements. To overcome this bottleneck, physical inductive biases must be incorporated in a manner that is highly cost-effective during both model training and inference.

Here, we introduce a highly efficient methodology to augment existing datasets with orbitally resolved,
semiempirical charge data. Our multitask learning framework demonstrates that training models to predict
computationally inexpensive GFN1-xTB orbital charges significantly improves the prediction of DFT total
energy and forces -- primary target properties of MLIPs. By forcing the network to learn a chemically rigorous latent space, GFN1-xTB orbital-charge supervision improves energy and force prediction over both energy-only and energy-and-force baselines, matches the best DFT atomic-charge target at a fraction of the cost, and unlike Mulliken charge supervision, never degrades accuracy. Because the semiempirical charges scale linearly $O(N)$ and the charge head is discarded at inference, the method adds negligible training overhead and none at inference.

While the present study highlights substantial performance gains in a relatively low-data regime, our dataset scaling analyses suggest that these improvements will persist and scale favorably into massive data regimes. The architecture utilized in this work is a lightweight 6.7 million parameter model. Translating these findings to scale up massive, frontier MLIP foundation models to achieve state-of-the-art accuracy presents a highly promising avenue for future research. Ultimately, leveraging semiempirical electronic structure data offers a scalable, physics-informed pathway to overcome current data bottlenecks, unlocking the next generation of highly accurate, sample-efficient chemical foundation models.

\vspace{-5pt}

\bibliographystyle{plainnat}
\bibliography{bibliography}

\newpage
\appendix

\section{Additional Results}\label{secA}

\begin{figure}[H]
    \centering
    \includegraphics[width=\linewidth]{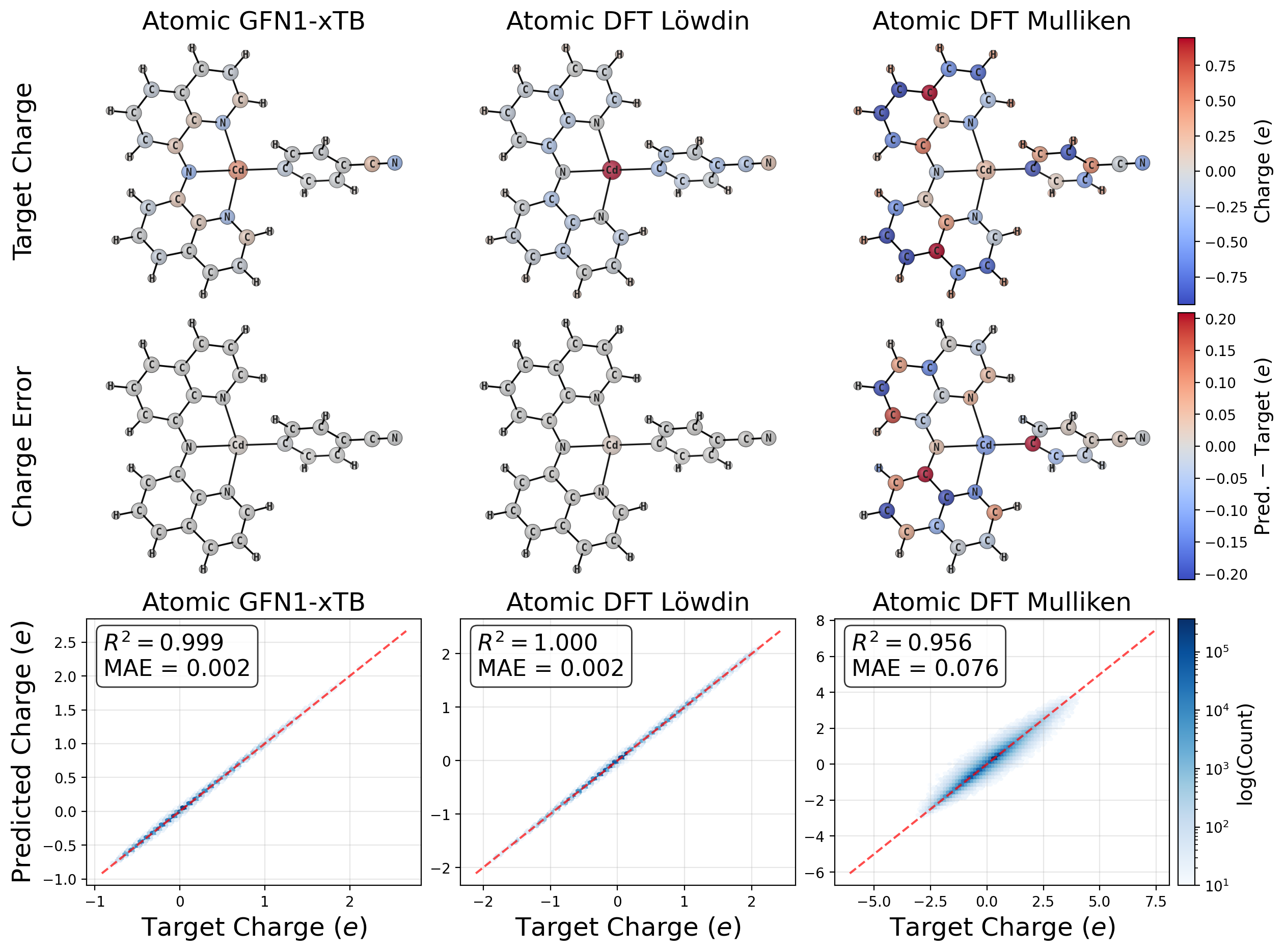}
    \caption{\textbf{Per-atom charge prediction for the three atomic-charge targets.}
    Columns compare models trained on atomic GFN1-xTB, DFT L\"owdin, and DFT Mulliken charges.
    Top row: a representative Cd complex colored by target atomic charge. The three
    partitioning schemes assign different charge distributions to the same geometry, with
    Mulliken exhibiting the most extreme charge separation. Middle row: the same molecule colored
    by atomic charge prediction error. Errors are near zero for GFN1-xTB and
    L\"owdin but large and structured for Mulliken. Bottom row: density plots of predicted
    vs.\ target charge over the test set. The model predicts GFN1-xTB and L\"owdin charges
    almost exactly, but fails to capture the highly polarized Mulliken charges. This reflects the well-known basis-set sensitivity of Mulliken population analysis, which produces a broad and less stable charge distribution, and may explain its harmful effect on energy accuracy reported in the main text.}
    \label{fig:atomic_charge_prediction}
\end{figure}

\begin{figure}[H]
    \centering
    \includegraphics[width=\linewidth]{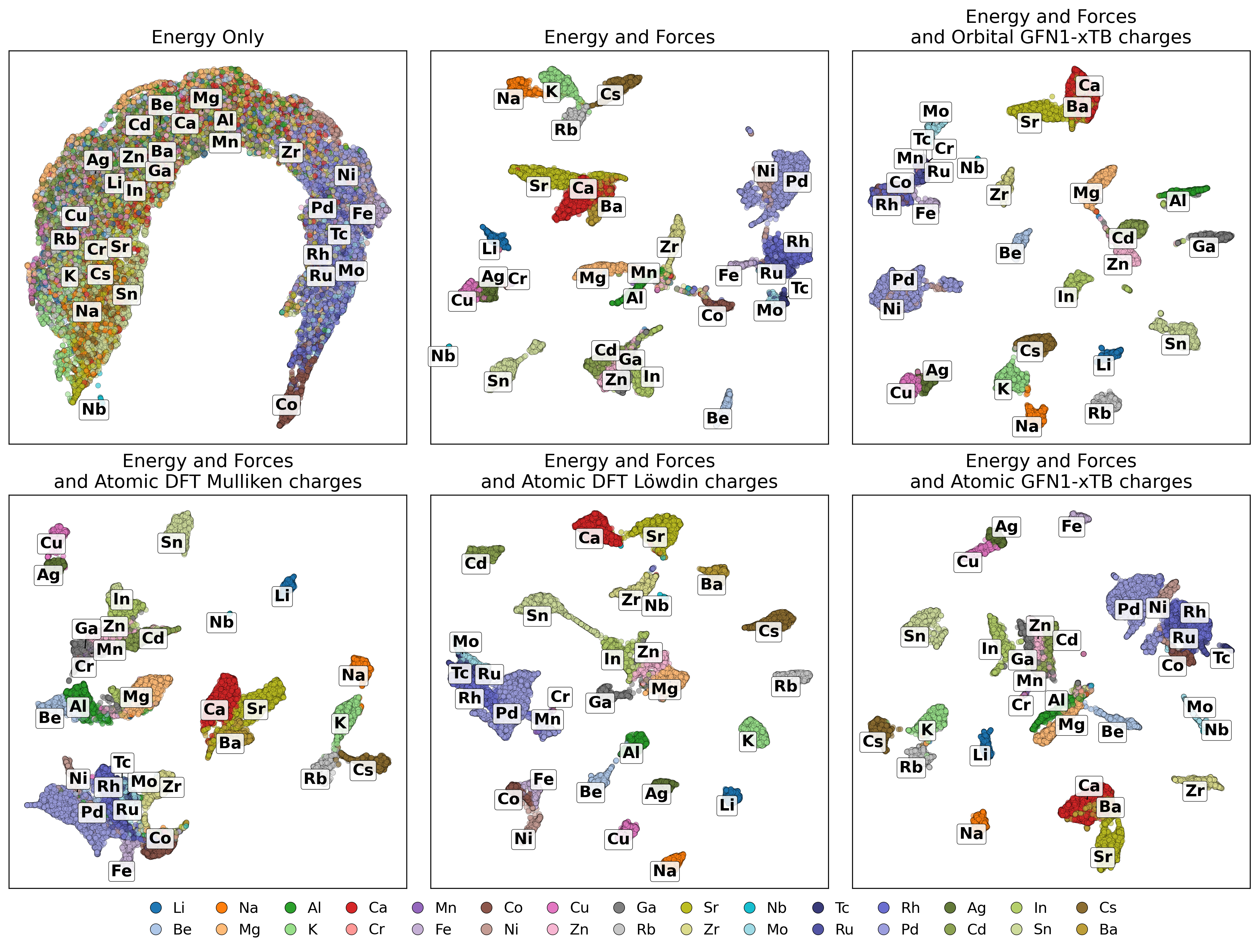}
    \caption{\textbf{UMAP projections of metal atomic embeddings across all six training setups.}
    Embeddings of 30 metallic elements extracted from the eSCN backbone after training,
    projected with UMAP and colored by element. Energy-only training (top left) produces a single
    diffuse manifold in which elements are not separated. Adding forces as a training task already induces partial clustering, and each charge-augmented target further sharpens it. Multitask training with orbital GFN1-xTB charges (top right) creates the most chemically accurate organization, spontaneously grouping elements of shared chemical character (e.g.\ Ag/Cu, Zn/Cd, Ca/Ba, Ni/Pd). The atomic-charge targets (bottom row) cluster elements to varying degrees but less cleanly than the orbital target.}
    \label{fig:umap_all_targets}
\end{figure}

This section provides supporting results for two claims made in the main text: that the choice
of charge-partitioning scheme, rather than its computational cost, governs how well a charge target
can be learned, and that orbital charge supervision creates the most chemically accurate latent
representation. Figure~\ref{fig:atomic_charge_prediction} examines per-atom charge prediction for
the three atomic-charge targets on a representative complex, showing that the network reproduces
GFN1-xTB and L\"owdin charges almost exactly while failing to accurately predict
Mulliken charges. Figure~\ref{fig:umap_all_targets} then compares the learned atomic embeddings
across all six training targets, extending the three-panel UMAP comparison of the main text and
confirming that the clustering behavior reported there is consistent across all training settings.

\section{Equivariant prediction of orbital charges}\label{secB}

In Section \ref{sec4}, we introduced the orbital charge definition and demonstrated how it can be predicted using the equivariant architecture of the eSCN model. In this section, we will derive orbital charge transformations under molecular coordinate rotations and show in detail how equivariant GNN models can efficiently capture the physics of orbital charges. For completeness, let's start by introducing all the necessary variables and concepts.

The transformation of the density matrix $\mathbf{P}$ and overlap matrix $\mathbf{S}$ under molecular coordinate rotations is captured by the rotation matrix $\mathbf{R}$ and the corresponding Wigner-D matrix {$\mathcal{D}(\mathbf{R})$}: 

\begin{equation}
        \mathbf{P'} = \mathcal{D} \mathbf{P} \mathcal{D}^T \quad \text{and} \quad \mathbf{S'} = \mathcal{D} \mathbf{S} \mathcal{D}^T
\end{equation}

The orbital populations $\bm{\nu}$ and charges $\mathbf{q}$ are computed from the density and overlap matrices, as well as the reference occupations $\bm\nu_0$ using element-wise multiplication: 
\begin{align}
    \nu_i &= \sum_{j=1}^{N_{orb}} (\mathbf{P} \circ \mathbf{S})_{ij} = \sum_{j=1}^{N_{orb}} P_{ij}S_{ij} \\
     \mathbf{q} &= \bm\nu_0 - \bm\nu
\end{align}

The updated orbital populations after rotation are computed as:
\begin{equation}
        \nu'_i = \sum_j (\mathbf{P'} \circ \mathbf{S'})_{ij} = \sum_j P'_{ij} S'_{ij}
\end{equation}

The elements of $\mathbf{P}$ and $\mathbf{S}$ transform as:
\begin{align}
    P'_{ij} &= \left( \mathcal{D} \mathbf{P} \mathcal{D}^T \right)_{ij} = \sum_{k,l} \mathcal{D}_{ik} P_{kl} (\mathcal{D}^T)_{lj} = \sum_{k,l} \mathcal{D}_{ik} P_{kl} \mathcal{D}_{jl} \\
    S'_{ij} &= \left( \mathcal{D} \mathbf{S} \mathcal{D}^T \right)_{ij} = \sum_{m,n} \mathcal{D}_{im} S_{mn} (\mathcal{D}^T)_{nj} = \sum_{m,n} \mathcal{D}_{im} S_{mn} \mathcal{D}_{jn}
\end{align}
Now substitute these into the expression for $\nu'_i$:
\begin{equation}
    \nu'_i = \sum_j \left( \sum_{k,l} D_{ik} P_{kl} D_{jl} \right) \left( \sum_{m,n} D_{im} S_{mn} D_{jn} \right)
\end{equation}
We can rearrange the summations, bringing the sum over $j$ to the inside.
\begin{equation}
    \nu'_i = \sum_{k,l,m,n} \mathcal{D}_{ik} \mathcal{D}_{im} P_{kl} S_{mn} \left( \sum_j \mathcal{D}_{jl} \mathcal{D}_{jn} \right)
\end{equation}
 Since $\mathcal{D}$ is an orthogonal matrix, we have $\mathcal{D}^T \mathcal{D} = \mathcal{I}$. In index notation:
\begin{equation}
    (\mathcal{D}^T \mathcal{D})_{ln} = \sum_j (\mathcal{D}^T)_{lj} \mathcal{D}_{jn} = \sum_j \mathcal{D}_{jl} \mathcal{D}_{jn} = \delta_{ln}
\end{equation}
where $\delta_{ln}$ is the Kronecker delta. Substituting this into our expression for $\nu'_i$:
\begin{equation}
    \nu'_i = \sum_{k,l,m,n} \mathcal{D}_{ik} \mathcal{D}_{im} P_{kl} S_{mn} \, \delta_{ln}
\end{equation}

The Kronecker delta collapses the sum over $n$, as the term is only non-zero when $n=l$. We replace every $n$ with $l$:
\begin{equation}
    \nu'_i = \sum_{k,l,m} \mathcal{D}_{ik} \mathcal{D}_{im} P_{kl} S_{ml}
\end{equation}
Let's regroup the terms to make the structure clearer:
\begin{equation}
    \nu'_i = \sum_{k,m} \mathcal{D}_{ik} \mathcal{D}_{im} \left( \sum_l P_{kl} S_{ml} \right)
\end{equation}

The term in the parentheses is the $(k,m)$-th element of the \emph{matrix product} $\mathbf{P}\mathbf{S}$ (since $S_{ml} = S_{lm}$ due to symmetry of the overlap matrix).
\begin{equation}
    \sum_l P_{kl} S_{ml} = \sum_l P_{kl} S_{lm} = (\mathbf{P}\mathbf{S})_{km}
\end{equation}

Substituting this back, we get:
\begin{equation}
    \nu'_i = \sum_{k,m} \mathcal{D}_{ik} (\mathbf{P}\mathbf{S})_{km} \mathcal{D}_{im}
\end{equation}

Now that we have the exact transformation rule of orbital populations, we can simplify it using the block-diagonal structure of the D-Matrix. Since $\mathcal{D}$ is block-diagonal by angular momentum, $\mathcal{D}_{ik}$ and $\mathcal{D}_{im}$ are only non-zero if $k$ and $m$ are in the same angular momentum block as $i$. Let this block be $\mathcal{B}(i)$. We can now write down the transformation rule of orbital populations, and therefore orbital charges, under molecular coordinate rotations:

\begin{equation}
        \nu'_i = \sum_{k,m \in \mathcal{B}(i)} \mathcal{D}_{ik} (\mathbf{P}\mathbf{S})_{km} \mathcal{D}_{im}
        \label{eq:A15}
\end{equation}

Our backbone model is an equivariant eSCN architecture that utilizes spherical harmonics. To efficiently predict orbital charges, we integrate the transformation rule from Equation \ref{eq:A15} into the charge prediction head of the model using the e3nn framework \citep{Geiger_Smidt_2022}. The model was implemented and trained using PyTorch, PyTorch Geometric, PyTorch Lightning, and Hydra \citep{paszke2017automatic, Fey_PyG_2_0_Scalable_2025, Falcon_PyTorch_Lightning_2019, Yadan2019Hydra}.

\section{Data generation and processing}\label{secC}

\begin{wrapfigure}[18]{r}{0.5 \textwidth}
    \centering
    \vspace{-5mm}
    \includegraphics[width=0.47 \textwidth]{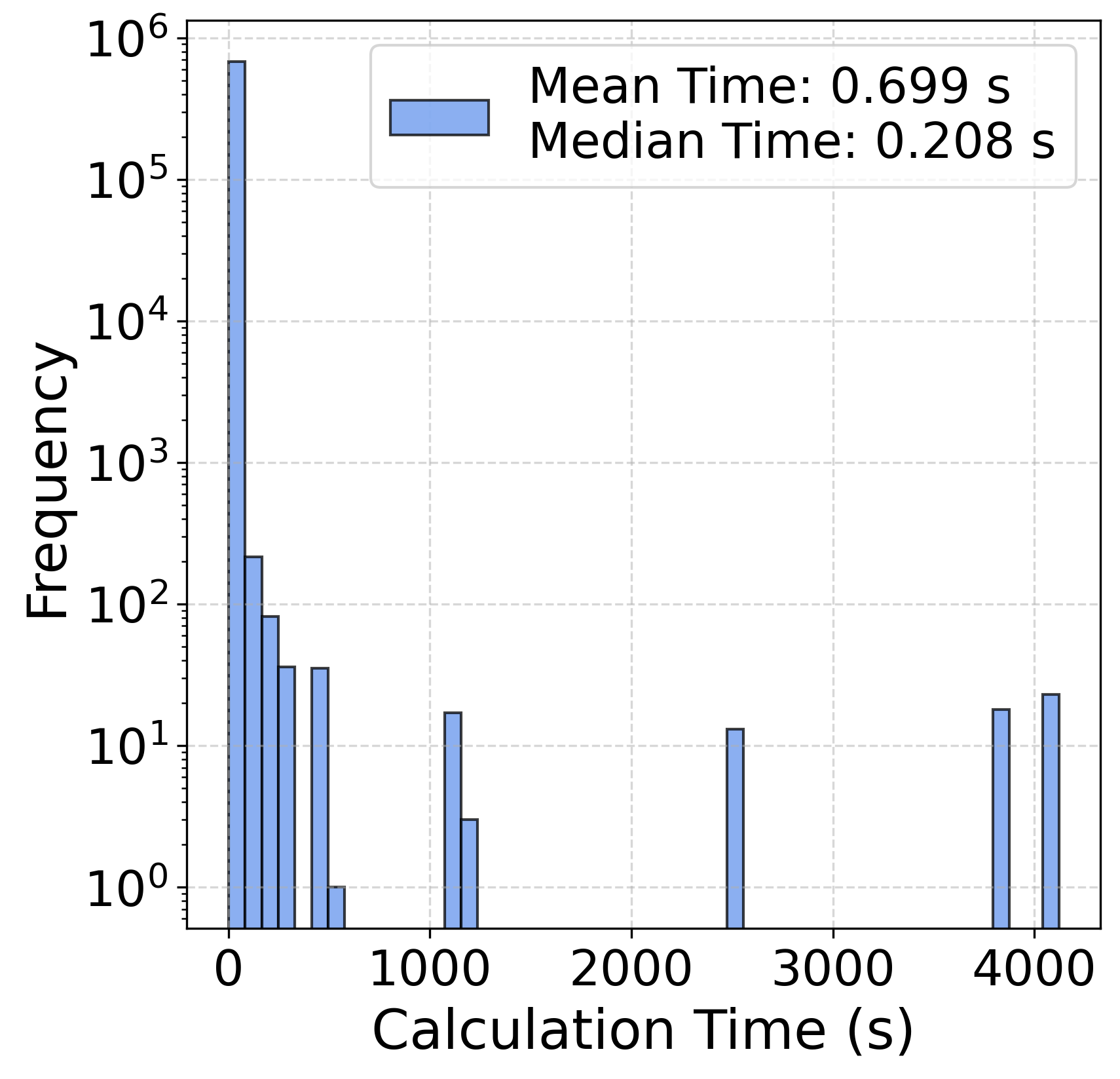}
    \vspace{-5mm}
    \caption{Distribution of the DXTB calculation times using one CPU core per molecule.}
    \label{fig:dxtb-calc-time}
\end{wrapfigure}

For this work, we extracted structures of neutral, singlet-state metal complexes from the OMol25 4M dataset. To compute orbital charges for the selected structures, we used the GFN1-xTB method implemented in the DXTB package\cite{friede_dxtb_2024}. The average time of computation per molecule is 0.7 seconds on a single CPU core (Fig. \ref{fig:dxtb-calc-time}).

The DXTB package allows selecting the target property for self-consistent field (SCF) cycle convergence (Fock matrix, charge, or potential) via \lstinline{SCP_MODE} parameter. We found that for organometallic systems, the results of DXTB calculations are not always consistent across all modes, even when convergence criteria are met, in some cases leading to unphysical results. We found that charge mode is the most reliable for organometallic systems. However, to remove potentially unreliable calculations, we filter out all structures where the final calculated orbital charges do not agree across all three modes. Furthermore, we removed several structures with maximum and minimum orbital charge values beyond $5\sigma$ of the distribution.

To compute GFN1-xTB atomic charge labels, we summed orbital charges on the respective atom of the structure. The DFT atomic charge labels from the Löwdin and Mulliken methods were obtained from the OMol25 dataset. To visualize molecular structures and atomic charge distributions in Fig. \ref{fig:atomic_charge_prediction}, we used the xyzrender code\cite{goodfellow_graph-based_2026}.

The DFT total energy labels were obtained from the OMol25 dataset. We computed reference-corrected total energies by fitting a linear model on the training set using element counts in each molecule as features. The linear model coefficients are presented in Table \ref{tab:coefficients}. The finalized dataset was converted into molecular graphs containing the 3D geometries alongside all target labels: DFT energies and forces, DFT Mulliken and L\"owdin atomic charges, and GFN1-xTB atomic
and orbital charges.

All models were trained on NVIDIA H100 GPUs. The total compute for the experiments reported in this work, spanning all six training settings, four nested dataset sizes, and three random seeds, was approximately 1,500 H100 GPU-hours.

\begin{table}[h!]
\centering
\small
\begin{tabular}{lcc|lcc|lcc}
\toprule
Sym & Z & Coeff & Sym & Z & Coeff & Sym & Z & Coeff \\
\midrule
H & 1 & -16.240 & Co & 27 & -37625.084 & Sb & 51 & -6537.447 \\
Li & 3 & -204.265 & Ni & 28 & -41041.040 & Te & 52 & -7293.723 \\
Be & 4 & -398.980 & Cu & 29 & -44637.789 & I & 53 & -8101.333 \\
B & 5 & -678.264 & Zn & 30 & -48414.137 & Cs & 55 & -545.329 \\
C & 6 & -1037.031 & Ga & 31 & -52373.654 & Ba & 56 & -691.333 \\
N & 7 & -1490.281 & Ge & 32 & -56515.645 & La & 57 & -857.862 \\
O & 8 & -2047.694 & As & 33 & -60838.778 & Ce & 58 & -12926.851 \\
F & 9 & -2717.553 & Se & 34 & -65346.863 & Yb & 70 & -31529.862 \\
Na & 11 & -4414.244 & Br & 35 & -70042.871 & Lu & 71 & -33619.560 \\
Mg & 12 & -5442.899 & Rb & 37 & -653.186 & Hf & 72 & -1306.258 \\
Al & 13 & -6596.852 & Sr & 38 & -833.443 & Ta & 73 & -1550.599 \\
Si & 14 & -7878.767 & Zr & 40 & -1280.331 & W & 74 & -1822.732 \\
P & 15 & -9289.050 & Nb & 41 & -1546.060 & Re & 75 & -2126.796 \\
S & 16 & -10835.703 & Mo & 42 & -1853.457 & Os & 76 & -2464.160 \\
Cl & 17 & -12522.600 & Tc & 43 & -2195.214 & Ir & 77 & -2836.620 \\
K & 19 & -16322.831 & Ru & 44 & -2578.888 & Pt & 78 & -3244.627 \\
Ca & 20 & -18437.128 & Rh & 45 & -3005.441 & Au & 79 & -3689.323 \\
Ti & 22 & -23116.356 & Pd & 46 & -3477.513 & Hg & 80 & -4167.575 \\
V & 23 & -25684.189 & Ag & 47 & -3996.714 & Tl & 81 & -4689.389 \\
Cr & 24 & -28416.584 & Cd & 48 & -4559.923 & Pb & 82 & -5242.944 \\
Mn & 25 & -31315.895 & In & 49 & -5170.207 & Bi & 83 & -5836.043 \\
Fe & 26 & -34384.770 & Sn & 50 & -5827.314 & Int & - & 1.461 \\
\bottomrule
\end{tabular}
\vspace{4mm}

\caption{Coefficients of the linear model used to correct the total energy reference (Symbol, Atomic Number Z, and Coefficient)}
\label{tab:coefficients}
\end{table}

\end{document}